\documentclass[12pt]{article}
\usepackage{amsmath}
\usepackage{amssymb}
\usepackage{bm}
\usepackage[dvips]{graphicx}

\begin{document}

\begin{titlepage}

 \setcounter{page}{0}

 \begin{flushright}
TIFR/TH/09-04 
 \end{flushright}

 \vskip 5mm
 \begin{center}
  {Modification of Gravitational Anomaly Method in Hawking Radiation}  

  \vskip 15mm

 {\large
 Takeshi Morita \footnote{\tt takeshi@theory.tifr.res.in}
  } 

  \vspace{5mm}

\small{\emph{ Department of Theoretical Physics, Tata Institute of Fundamental Research,}} \\
\small{\emph{Homi Bhabha Rd, Mumbai 400005, India}} \\ [1mm]
  
 \end{center}

\vskip 30mm 

 \centerline{{\bf{Abstract}}} 

\vskip 3mm
We discuss an ambiguity of the derivation of the Hawking radiation through the gravitational anomaly method and propose modifications of this method such that it reproduces the correct thermal fluxes.
In this modified gravitational anomaly method, we employ the two-dimensional conformal field theory technique.
\end{titlepage}

\newpage

\section{Introduction}
\setlength{\baselineskip}{7mm}
Hawking radiation from black holes is one of the most important effects in black hole thermodynamics and the quantum effect of gravity \cite{Hawking:1974rv, Hawking:1974sw}.
There are several derivations of Hawking radiation and recently one interesting method was proposed by Robinson and Wilczek \cite{Robinson:2005pd}.  
They considered the effective chiral theory near the horizon and showed that the gravitational anomaly \cite{AlvarezGaume:1983ig} in this effective theory causes the energy flux at the radial infinity which can be identified as Hawking radiation.
This effective chiral theory would be related to the effective theory on the membrane in the membrane paradigm \cite{Parikh:1997ma,Parikh:1999mf,Iqbal:2008by,Thorne:1986iy} and thus this derivation suggests the association between the Hawking effect and the membrane paradigm. 
This derivation would also connect the Hawking effect with some phenomena in condensed matter physics.

This new interpretation of the Hawking effect was modified by Iso, Umetsu and Wilczek \cite{Iso:2006wa,Iso:2006ut}.
Furthermore this method was simplified by using the covariant currents \cite{Banerjee:2007qs} and the spectra of the thermal distribution functions were also reproduced by considering the higher-spin currents \cite{Iso:2007kt,Iso:2007hd,Iso:2007nc,Iso:2007nf,Bonora:2008nk,Bonora:2008he}.
Further developments and the generalization to various black holes were also shown by many authors \cite{Murata:2006pt,Vagenas:2006qb,Setare:2006hq,Xu:2006tq,Iso:2006xj, Jiang:2007wj, Das:2007ru,Jiang:2007pe, Banerjee:2007uc,Huang:2007ed,Gangopadhyay:2007hr,  Iso:2008sq, Umetsu:2008cm,  Shirasaka:2008yg, Banerjee:2008az,Akhmedova:2008au,
Banerjee:2008wq,Morita:2008qn,Papantonopoulos:2008wp,Wei:2009kg}.

However there is one problem with this derivation.
Hirata and Shirasaka \cite{Shirasaka:2008yg} found a constant of integration which had not been considered in the calculation of the gravitational anomaly method.
We will show that the flux is not fixed owing to this constant.

The purpose of this study is to modify the gravitational anomaly method such that it reproduces the correct fluxes.
We will show that in the case of the U(1) current we can derive the flux by considering the chiral current and in the case of the energy-momentum tensor we can derive it by considering the trace anomaly.
In these derivations, we will employ the calculation of the fluxes based on the two-dimensional conformal field theory technique \cite{Christensen:1977jc, Iso:2006ut,Iso:2007hd}.

In section \ref{sec ambiguity}, we show the ambiguity in the gravitational anomaly method and we argue for the modifications in section \ref{Sec modification}.
In section \ref{Sec energy}, we apply this modification to the derivation of the energy flux. 
Section \ref{Conclusion} contains conclusions and discussions.
In Appendix \ref{App-RN}, we summarize the basics of Reissner-Nordstr\"om black holes.

\section{Ambiguity in Gravitational Anomaly Method}
\label{sec ambiguity}

We show that the gravitational anomaly method has an ambiguity and discuss the problem with it.
We investigate the derivation of the flux of the U(1) current from a $4$-dimensional Reissner-Nordstr\"om black hole as an example. 
It will be possible to generalize this argument to other currents and black holes.

First we attempt to derive the flux through the gravitational anomaly method \cite{Iso:2006wa,Banerjee:2007qs}.
We consider a matter field in the Reissner-Nordstr\"om black hole background.
(See Appendix \ref{App-RN} for the Reissner-Nordstr\"om solution.)
It is known that the matter field near the horizon can be effectively described as massless fields in two dimensions $(t,r_*)$.  
Then the covariant U(1) current $J^\mu$ satisfies the two-dimensional conservation law \cite{Iso:2007nc}
\begin{align}
 \nabla_\mu J^{\mu}
& = -\frac{(c_R-c_L)}{2}\frac{ e^2}{2\pi} \epsilon^{\mu\nu}  F_{\mu\nu}
\label{U(1) conservation}. 
\end{align}
Here $e$ is the electric charge of the matter and $F_{\mu\nu}$ is the background field strength.
$\epsilon^{\mu\nu}$ is  the covariant antisymmetric tensor.
$c_L$ and $c_R$ are the central charges of the left and right modes, which correspond to the in-going and out-going modes in the black hole background, and $c_L=c_R=1$ ($c_L=c_R=1/2$) if the matter is a real
 boson (fermion).
Note that the central charge of a charged field is twice that of a real field, since it is a complex field.
Thus the right-hand side of this equation would vanish in all these cases.

In the gravitational anomaly method \cite{Robinson:2005pd, Iso:2006wa}, the in-going modes, which are classically irrelevant to physics outside the horizon, are eliminated near the horizon and we divide the outside of the horizon into two: the near horizon region $(r_+<r<r_++\epsilon)$ and the out region $(r_++\epsilon<r<\infty)$\footnote{Since the two-dimensional description is effective near the horizon only, we cannot take $r$ a large value. 
However we use this description even if $r \gg r_+$. 
It is known that the fluxes which are derived through this approximation are equivalent to the $4$-dimensional fluxes without the gray body factor. }.
Here $r_+$ is the radius of the outer horizon and $\epsilon$ is an appropriately small parameter.
In the near horizon region, the effective theory is chiral since the in-going modes do not contribute.
It means that the current in this region satisfies  the conservation equation (\ref{U(1) conservation}) with $c_L=0$ and thus the current is anomalous.
On the other hand, in the out region, the effective theory is still non-chiral ($c_L=c_R$).
Then the U(1) current can be described as
\begin{align}
J^\mu=J^\mu_{(O)}\Theta_+(r)+  J_{(H)}^{\mu}  H(r),
\end{align}
where we have employed step function $\Theta_+(r)=\Theta(r-(r_++\epsilon))$ and $H(r)=1-\Theta_+(r)$.
$J^\mu_{(O)}$ denotes the current in the out region and $J^\mu_{(H)}$ denotes the current in the near horizon region.
These currents satisfy 
\begin{align}
 \nabla_\mu J_{(O)}^{\mu}
& = 0,
\label{conservation JO}
\\
 \nabla_\mu J_{(H)}^{\mu}
& = -c_R\frac{ e^2}{4\pi} \epsilon^{\mu\nu}  F_{\mu\nu},
\label{conservation JH}
\end{align}
respectively.
Now we consider the total current $J^\mu_{(total)}$ including the contribution from the near horizon in-going modes.
This current should satisfy the conservation equation (\ref{U(1) conservation}) with $c_L=c_R$ and it can be described as
\begin{align}
J_{(total)}^\mu=J^\mu+K^\mu  H(r)+j_{(total)}^\mu.
\label{total current}
\end{align}
Here $j_{(total)}^\mu$ is a possible additional current which satisfies $\nabla_\mu j_{(total)}^\mu=0$ and $K^\mu$ is the contribution of the in-going modes which satisfies
\begin{align}
 \nabla_\mu K^{\mu}
& = c_L\frac{ e^2}{4\pi} \epsilon^{\mu\nu}  F_{\mu\nu} .
\label{conservation K}
\end{align}
In addition, these currents should satisfy
\begin{align} 
J_{(O)}^{\mu}=J_{(H)}^{\mu}+K^{\mu}
\label{coincident}
\end{align} 
 at $r=r_++\epsilon$ such that $\nabla_\mu J^\mu_{(total)}=0$.
Since the black hole background is static, the current does not depend on time. 
Then we can solve the equation (\ref{conservation JO}), (\ref{conservation JH}) and (\ref{conservation K}) by integrating them,\footnote{We evaluate these equations by using the Schwarzschild coordinates. 
However these coordinates are not appropriate for the calculation of the Hawking effect and we should employ the tortoise coordinate. However we can obtain the same result and we use the Schwarzschild coordinates since the expressions of equations are simpler. }
\begin{align}
 J_{(O)}^{r}
& = j_{(O)}^r,\\
 J_{(H)}^{r}
& = c_R\frac{ e^2}{2\pi}  A_t (r)+ j_{(H)}^r, \label{solution J_H}\\
 K^{r}
& = -c_L\frac{ e^2}{2\pi}  A_t (r)+ k^r,
\end{align}
where we have used $\epsilon^{rt}=-1$.
Here $j_{(O)}^r,j_{(H)}^r$ and $k^r$ are integral constants.
Especially $j_{(O)}^r$ will correspond to the flux which is observed at the infinity.
The existence of the integral constant $k^r$ was pointed out by Hirata and Shirasaka in \cite{Shirasaka:2008yg} but they took $k^r=0$ in their calculation.
This constant will cause the ambiguity as we will argue later.

\cite{Iso:2006wa, Banerjee:2007qs} impose the following two conditions:
\begin{align} 
J^r=0 \quad \text{at}~ r=r_+, \quad j_{(total)}^r=0.
\label{old condition}
\end{align} 
These conditions were supposed to correspond to the Unruh vacuum \cite{Iso:2006ut}, which we will discuss in the next section.
Then the integral constants satisfy,
\begin{align}
j_{(H)}^r&=-c_R\frac{ e^2}{2\pi}  A_t(r_+),\\
j_{(O)}^r&= -c_R\frac{ e^2}{2\pi}  A_t(r_+)+k^r,
\end{align} 
where we have considered the equation (\ref{coincident}). 
Thus we obtain $J^r=-c_R e^2  A_t(r_+)/2\pi+k^r$ at the infinity and the flux is not fixed.
The correct flux, which is expected in the black hole thermodynamics, is $J^r=-c_R e^2  A_t(r_+)/2\pi$ and it is obvious that $k^r$ causes the ambiguity.
Surely we can remove this ambiguity by imposing the additional condition $k^r=0$ as in \cite{Shirasaka:2008yg}.
However the physical meaning of this condition is not clear.
We can find a similar ambiguity in the derivation of the energy flux also.

\section{Modification of Gravitational Anomaly Method}
\label{Sec modification}

We discuss the modifications of the gravitational anomaly method by considering the chiral current $J^{5\mu}$.
We can solve the anomalous conservation equation of $J^{5\mu}$ in the $(t,r)$ coordinates as we calculated in the previous section, but the light-cone coordinate $(u,v)$ (\ref{light-cone}) are much useful and we will use them.

Before considering the gravitational anomaly method, we review the derivation of the flux based on the two-dimensional conformal field theory technique \cite{Christensen:1977jc, Iso:2006ut,Iso:2007hd} since this derivation illuminates our problem.

The two-dimensional chiral current is defined by $J^{5\mu}=\epsilon^{\mu\nu}J_{\nu}$, where  the covariant antisymmetric tensor is $\epsilon^{uv}=2e^{-\varphi}$ in the $(u,v)$ coordinates and $\varphi$ is the background metric (\ref{light-cone}).
$J^{5\mu}$ satisfies the anomalous conservation equation (the chiral anomaly) \cite{Iso:2007nc},
\begin{align}
 \nabla_\mu J^{5\mu}
& = \frac{(c_L+c_R)}{2}\frac{ e^2}{2\pi} \epsilon^{\mu\nu}  F_{\mu\nu}.
\label{chiral conservation}
\end{align}
By taking the Lorentz gauge $\partial_u A_v+\partial_v A_u=0$ for the background gauge field, we can solve this equation and (\ref{U(1) conservation}) as
\begin{align}
 J_{u}=j_u+c_R\frac{ e^2}{\pi}A_u, \quad
 J_{v}= j_v +c_L\frac{ e^2}{\pi}A_{v}.
\label{uv current }
\end{align}
Here $j_u$ and $j_v$ are integral constants.
Strictly speaking, $j_u$ and $j_v$ should be holomorphic functions with respect to $u$ and $v$ respectively. 
However since the background is time independent, we can take them as constants.
Note that $J_{u}$ $(J_v)$ corresponds to the out-going (in-going) current.

We can derive the fluxes by imposing the following boundary conditions: 
\begin{enumerate}
  \item Regularity condition: $J_u=0$ at the horizon.
  \item No in-going flux at the infinity: $J_v=0$ at $r=\infty$.
\end{enumerate}
The first condition means that the free falling observer does not observe the singular flux at the horizon.
It is known that these conditions are corresponding to the Unruh vacuum.
(Note that the Boulware vacuum corresponds to the condition $J_u=J_v=0$ at $r=\infty$ and the Hartle-Hawking vacuum corresponds to $J_u=J_v=0$ at the horizon \cite{Birrell:1982ix}.)
Then the integral constants are fixed as
\begin{align} 
j_u =-c_R\frac{ e^2}{\pi}A_{u}(r_+), \quad j_v=0.
\end{align} 
Thus we obtain the correct flux at the infinity,
\begin{align} 
J^r(r\rightarrow \infty)&=J_u(r\rightarrow \infty)-J_v(r\rightarrow \infty)\nonumber \\
&=-c_R\frac{ e^2}{2\pi}A_{t}(r_+).
\label{J^r flux}
\end{align} 
This is the derivation of the flux associated with the U(1) current through the conformal field theory technique.

Now we discuss the gravitational anomaly method by considering this derivation.
As we argued in the previous section, we take $c_L=0$ in the near horizon region.
It implies that the in-going current (\ref{uv current }) is modified as
\begin{align}
J_{v}&=J_{(O) v}\Theta_+(r)+ J_{(H) v} H(r),\\
J_{(total)v}&=J_v+ K_v H(r)+j_{(total)v}, \\
J_{(O) v}&=j_{(O) v}+c_L\frac{ e^2}{\pi}A_v, \quad
 J_{(H) v}=j_{(H) v}, \\
K_v& = k_v +c_L \frac{e^2}{\pi}A_v.
\end{align}
Here $j_{(total)v} ,j_{(O) v},j_{(H) v}$ and $ k_v$ are integral constants.
$J_{(H) v}$ is the in-going current in the near horizon and $J_{(O) v}$ is in the out region. 
$K_v$ denotes the contribution of the in-going modes.
Similarly the out-going current becomes
\begin{align}
J_{u}&=J_{(O) u}\Theta_+(r)+ J_{(H) u} H(r),\\
J_{(total)u}&=J_u+ K_u H(r)+j_{(total)u}, \\
J_{(O) u}&=j_{(O) u}+c_R\frac{ e^2}{\pi}A_u, \quad
 J_{(H) u}=j_{(H) u}+c_R \frac{e^2}{\pi}A_u, \\
K_u& = k_u .
\end{align}
Here $j_{(total)u} ,j_{(O) u},j_{(H) u}$ and $ k_u$ are integral constants.
Then it is obvious that $J_{(H)u}$ and $J_{(H)v}$ satisfy the equation (\ref{U(1) conservation}) and (\ref{chiral conservation}) with $c_L=0$ in the near horizon region.
By considering the conservation equations of the total currents, the integral constants satisfy
\begin{align}
j_{(O) v}= j_{(H) v}+k_v, \quad
j_{(O) u}= j_{(H) u}+k_u,
\end{align}
as in (\ref{coincident}).
The relations between the integral constants in the previous section and in this section are as follows:
\begin{align}
j_{(O) }^r&=j_{(O) u}-j_{(O) v},\quad j_{(H) }^r=j_{(H) u}-j_{(H) v}, \nonumber \\
j_{(total)}^r&=j_{(total)u}-j_{(total)v}, \quad k^r=k_u-k_v.
\end{align}

In order to derive the flux, we consider the boundary conditions for the currents.
In \cite{Iso:2006wa,Iso:2006ut}, since they did not consider $k_\mu$, other constants were supposed to satisfy $j_{(O)u}=j_{(H)u}$ and $j_{(O)v}=j_{(H)v}$.  In this case, $j_{(O)u}$ and $j_{(O)v}$ are not distinguishable from $j_{(total)u}$ and $j_{(total)v}$ respectively and they took them as $j^r_{(O)}=j_{(O)u}$, $j^r_{(total)}=-j_{(total)v}$ and $j_{(total)u}=j_{(O)v}=j_{(H)v}=0$ in our notation.
Then the conditions in (\ref{old condition}) are corresponding to the Unruh vacuum.
However we now consider $k_\mu$ and these conditions are not valid.

We impose the following conditions for the currents instead of the condition (\ref{old condition}).
First we take
\begin{align} 
j_{(total)u}=j_{(total)v}=0.
\label{constraint total}
\end{align} 
The meaning of these conditions is as follows.
In the out region, $J_{(O)}^\mu$ associates with the excitation of the matter field.
The observer at the infinity observes this excitation and thus the observable must be $J_{(O)}^\mu$ only.
Thus we take these conditions and ignore $j_{(total) \mu}$ in our derivation.

Secondly we take
\begin{align} 
k_u=0.
\label{constraint in-going}
\end{align} 
This condition means that $K_u$ does not contribute to the out-going flux since $K_u$ is the contribution from the in-going modes.

In addition to these conditions, we impose the boundary conditions corresponding to the Unruh vacuum:
\begin{align}
J_{u}=0 \quad \text{at}~ r=r_+, \quad J_{v}=0 \quad \text{at}~r=\infty .
\label{boundary Unruh}
\end{align} 
Then we obtain the flux at the infinity,
\begin{align} 
j_{(O) }^r=j_{(O) u}=j_{(H) u}=-c_R\frac{ e^2}{\pi}A_u(r_+).
\end{align} 
This equation implies that the origin of the flux at the infinity is $j_{(H) u}$ in the near horizon chiral theory.
Thus the Hawking effect can be regarded as the contribution of the near horizon anomalies.
Note that $k^r$, which causes the ambiguity of the flux in the previous section,  has not been fixed.
Even though we could obtain the correct flux since we have derived the in-going and out-going currents at the infinity separately.

Here we summarize the derivation of the modified gravitational anomaly method.
\begin{enumerate}
  \item Divide the outside of the horizon into two and take $c_L=0$ in the near horizon and $c_L=c_R$ in the out region.
  \item Solve the conservation equations (\ref{U(1) conservation}) and (\ref{chiral conservation}) in each region.
  \item Impose the conditions (\ref{constraint total}) and (\ref{constraint in-going}) on the integral constants.
  \item Impose the boundary condition (\ref{boundary Unruh}) corresponding to the Unruh vacuum.
\end{enumerate}
Through this procedure, we can derive the flux from the anomalies in the near horizon.
In addition, we can easily show that if we impose the boundary conditions corresponding to the Boulware vacuum or the Hartle-Hawking vacuum instead of the Unruh vacuum,  the correct flux can be derived through the same procedure\footnote{In the case of the Hartle-Hawking vacuum, the boundary condition of the in-going modes at the horizon is imposed on the total current $J_{(total) v}$ rather than $J_{v}$ in order to reproduce the correct flux. It implies that the effective chiral theory near the horizon is not essential in this case.}.

\section{Derivation of Energy Flux through Modified Gravitational Anomaly Method}
\label{Sec energy}

In this section, we consider the derivation of the energy flux through the modified gravitational anomaly method.
As in the derivation of the U(1) current, the anomalous conservation equation of the energy-momentum tensor is not sufficient to derive the energy flux at the infinity and we need to consider the trace anomaly equation also.
These equations are given by, 
\begin{align} 
\nabla^\mu T_{\mu\nu}=& 
 F_{\mu\nu} J^\mu 
 - \frac{c_R-c_L}{96\pi} \epsilon_{\mu \nu} \nabla^\mu  R,  \\
{T^\mu}_{\mu}=&\frac{c_L+c_R}{48\pi}R,  
\end{align} 
where $R$ denotes the two-dimensional Ricci scalar \cite{Iso:2007nc}.
We can solve these equations as,
\begin{align} 
 T_{uu}  &= t_{uu}
  + 2 A_u j_{u}  + \frac{c_R e^2}{\pi} A_u^2
+ \frac{c_R}{24\pi}
  \left(\partial_u^2 \varphi - \frac{1}{2}(\partial_u \varphi)^2\right),\\
  T_{vv}  &= t_{vv}
  + 2 A_v j_{v}  + \frac{c_L e^2}{\pi} A_v^2
+ \frac{c_L}{24\pi}
  \left(\partial_v^2 \varphi - \frac{1}{2}(\partial_v \varphi)^2\right).
\end{align} 
Here $t_{uu}$ and $t_{vv}$ are integral constants and $\varphi$ is the background gravity (\ref{light-cone}).

Now we regard the near horizon theory as chiral and divide the outside of the horizon.
Then we can obtain the currents.
The in-going current becomes
\begin{align} 
T_{vv}=&T_{(O)vv}\Theta_+(r)+ T_{(H) vv} H(r)
, \\
T_{(total)vv}=&T_{vv}+K_{vv}  H(r)+t_{(total)vv},\\
  T_{(O)vv}  =& t_{(O)vv}
  + 2 A_v j_{(O)v}  + \frac{c_L e^2}{\pi} A_v^2 + \frac{c_L}{24\pi}
  \left(\partial_v^2 \varphi - \frac{1}{2}(\partial_v \varphi)^2\right),\\
    T_{(H)vv}  =& t_{(H)vv}
  + 2 A_v j_{(H)v}  ,\\
  K_{vv}  =& k_{vv}
  + 2 A_v k_{v}  + \frac{c_L e^2}{\pi} A_v^2 + \frac{c_L}{24\pi}
  \left(\partial_v^2 \varphi - \frac{1}{2}(\partial_v \varphi)^2\right),
\end{align} 
and the out-going current becomes
\begin{align} 
T_{uu}=&T_{(O)uu}\Theta_+(r)+ T_{(H) uu} H(r)
, \\
T_{(total)uu}=&T_{uu}+K_{uu}  H(r)+t_{(total)uu},\\
  T_{(O)uu}  =& t_{(O)uu}
  + 2 A_u j_{(O)u}  + \frac{c_R e^2}{\pi} A_u^2 + \frac{c_R}{24\pi}
  \left(\partial_u^2 \varphi - \frac{1}{2}(\partial_u \varphi)^2\right),\\
  T_{(H)uu}  =& t_{(H)uu}
  + 2 A_u j_{(H)u}  + \frac{c_R e^2}{\pi} A_u^2 + \frac{c_R}{24\pi}
  \left(\partial_u^2 \varphi - \frac{1}{2}(\partial_u \varphi)^2\right),\\
    K_{uu}  =& k_{uu}
  + 2 A_u k_{u}  .
 \end{align} 
Here $t_{(total)vv}$, $t_{(total)uu}$, $t_{(O)vv}$, $t_{(O)uu}$, $t_{(H)vv}$, $t_{(H)uu}$, $k_{vv}$ and $k_{uu}$ are integral constants.
These constants satisfy $t_{(O)vv}=t_{(H)vv}+k_{vv}$ and $t_{(O)uu}=t_{(H)uu}+k_{uu}$.

By imposing the condition $t_{(total)uu}=t_{(total)vv}=0$ and $k_{uu}=0$ and the boundary conditions corresponding to the Unruh vacuum, we obtain
\begin{align}
t_{(O)uu}= 
t_{(H)uu}=&-2 A_u (r_+)j_{(H)u}  - \frac{c_R e^2}{\pi} A_u^2(r_+) - \frac{c_R}{24\pi} \left(\partial_u^2 \varphi (r_+)- \frac{1}{2}(\partial_u \varphi(r_+))^2\right) \nonumber \\
  =&\frac{c_R}{192\pi}(f'(r_+))^2+\frac{c_R e^2}{\pi}A_u^2(r_+),\\
t_{(O)vv}=&0.
\end{align} 
Then the energy flux at the infinity is given by
\begin{align} 
{T^r}_t(r\rightarrow \infty)=&T_{uu}(r\rightarrow \infty)-T_{vv}(r\rightarrow \infty) \nonumber \\
=&\frac{c_R}{192\pi}(f'(r_+))^2+\frac{c_R e^2}{\pi}A_u^2(r_+).
\end{align} 
This result is coincident with the known result \cite{Iso:2006wa}.

\section{Conclusions and Discussions}
\label{Conclusion}

In this Letter, we have discussed the problem with the ambiguity of the gravitational anomaly method.
We have shown that, by considering the chiral current and the trace anomaly, the correct fluxes can be derived.
Thus we can interpret the origin of the fluxes as the anomalies in the near horizon.

Although we can derive the flux by using the conformal field theory technique without employing the near horizon chiral theory as we showed in section \ref{Sec modification}, the gravitational anomaly method is attractive since it would relate the Hawking effect to the membrane paradigm and condensed matter physics.

Another derivation of the Hawking effect associated with the gravitational anomaly method was proposed by Banerjee et al \cite{Banerjee:2008az,Banerjee:2008wq}.
They omitted the separation of the outside of the horizon and applied the anomaly equation (\ref{conservation JH}) to the theory in the whole region of the outside.
If we impose the condition $J^r_{(H)}=0$ at the horizon,
we can obtain the flux $j^r_{(H)}$ and they interpreted that this flux is the Hawking radiation observed at the infinity.
If we admit this derivation, the ambiguity which we have discussed in this article does not exist.
However this derivation is physically not correct, since the theory in the region apart from the horizon is not  anomalous and we cannot use (\ref{conservation JH}) in this region.
In addition, the expectation value of the current $J^r_{(H)}$ is not coincident with the correct current at finite $r$ because of the existence of the anomalous term $e^2 A_t(r)/2\pi$ in (\ref{solution J_H}).
Thus we avoided using the derivation \cite{Banerjee:2008az,Banerjee:2008wq} in this article.

\paragraph{Acknowledgements }
I would like to acknowledge useful discussions with T. Hirata, S. Iso,  A. Shirasaka and H. Umetsu.
I would also especially like to thank G. Mandal for useful discussions and for several detailed comments on the manuscript.

\appendix

\section{Reissner-Nordstr\"om black hole}
\setcounter{equation}{0}
\label{App-RN}

We summarize the basics of Reissner-Nordstr\"om black holes.  The metric and
the gauge potential of Reissner-Nordstr\"om black holes with mass $M$ and
charge $Q$ are given by
\begin{align} 
 ds^2 &= f(r)dt^2-\frac{1}{f(r)}dr^2-r^2d\Omega_2^2,\\
 A_t &= -\frac{Q}{r}, \label{bg-gauge}
\end{align} 
where 
\begin{align} 
 f(r) = 1-\frac{2M}{r}+\frac{Q^2}{r^2} = \frac{(r-r_+)(r-r_-)}{r^2}
\end{align} 
 and the radius of outer (inner) horizon $r_\pm$ is given by
 \begin{align} 
 r_\pm = M \pm \sqrt{M^2-Q^2}.
\end{align} 
It is useful to define the  tortoise coordinate 
by solving $  dr_* = dr/f $  as
\begin{align} 
  r_* = r + \frac{1}{2\kappa_+}\ln \frac{|r-r_+|}{r_+}
  + \frac{1}{2\kappa_-}\ln \frac{|r-r_-|}{r_-}.
\end{align} 
  Here the surface gravity at $r_\pm$ is given by
 \begin{align} 
  \kappa_\pm = \frac{1}{2}f'(r_\pm) = \frac{r_\pm - r_\mp}{2r_\pm^2}.
\end{align} 
We define the light-cone coordinates, 
$u=t-r_*$ and $v=t+r_*$. $u(v)$ are the 
out-going (in-going) coordinates and the metric 
in these coordinates becomes as
\begin{align} 
 ds^2 = f(dt^2 - dr_*^2) - r^2d\Omega^2 
 = fdudv - r^2d\Omega^2. 
 \label{uv}
\end{align} 
If we restrict to see the two-dimensional $(r,t)$ section,
both of these coordinates (\ref{uv}), have the forms of 
the conformal gauge
\begin{align} 
 ds^2 = e^{\varphi(u,v)} dudv,
\label{light-cone}
\end{align} 
where $\varphi=\log f$.
In this coordinate, the gauge potential becomes $A_u=A_v=A_t/2$.


\bibliographystyle{utphys}
\bibliography{mRW}

\end{document}